\documentclass[referee]{aa}
\usepackage{psfig}
\usepackage{graphicx}
\def\M21{\hbox{Mrk\,421} }
\def\xmm{\hbox{$XMM-Newton$} }
\def\etal{et al.\/ }
\def\ie{i.e., }

\def\ltsima{$\; \buildrel < \over \sim \;$}
\def\simlt{\lower.5ex\hbox{\ltsima}}            
\def\gtsima{$\; \buildrel > \over \sim \;$}
\def\simgt{\lower.5ex\hbox{\gtsima}}            
\def\eg{e.g., }
\newcommand{\eqref}[1]{(\ref{#1})}

\begin{document}

\title{Synthetic X-ray light curves of BL Lacs from
       relativistic hydrodynamic simulations}
\author{P. Mimica\inst{1}, M.A. Aloy\inst{1}, E. M\"uller\inst{1} and
W. Brinkmann\inst{2}}
\offprints{PM, e-mail: pere@mpa-garching.mpg.de}
\institute{Max--Planck--Institut f\"ur Astropysik,
           Postfach 1312, D-85741 Garching, FRG 
\and       Max-Planck-Institut f\"ur extraterrestrische Physik,
           Postfach 1603, D-85740 Garching, Germany, FRG}
\date{Received ?; accepted ?}
 \abstract{
 We present the results of relativistic hydrodynamic simulations of
 the collision of two dense shells in a uniform external medium, as
 envisaged in the internal shock model for BL Lac jets.  The
 non-thermal radiation produced by highly energetic electrons injected
 at the relativistic shocks is computed following their temporal and
 spatial evolution. The acceleration of electrons at the relativistic
 shocks is parametrized using two different models and the
 corresponding X-ray light curves are computed. We find that the
 interaction time scale of the two shells is influenced by an
 interaction with the external medium. For the chosen parameter sets,
 the efficiency of the collision in converting dissipated kinetic
 energy into the observed X-ray radiation is of the order of one
 percent.
  }
\titlerunning{Computing X-ray light curves of blazars}
\authorrunning{Mimica et. al}
\maketitle

\section{Introduction}
\smallskip BL Lac objects are thought to be dominated by relativistic
jets seen at small angles to the line of sight (\cite{UP95} 1995), and
their remarkably featureless radio-through-X-ray spectra are well
fitted by inhomogeneous jet models (\cite{BMU87} 1987).  As the
measured spectra can be reproduced by models with widely different
assumptions the structure of the relativistic jets remains largely
unknown. Only the analysis of the temporal variations of the emission,
and combined spectral and temporal information can considerably
constrain the jet physics.  Time scales of the observed light curves
are related to the crossing time of the emission regions which depend
on wavelength and/or the time scales of micro-physical processes like
particle acceleration and radiative losses.  The measured time lags
between the light curves at different energies as well as spectral
changes during intensity variations allow to probe the microphysics of
particle acceleration and radiation in the jet.
  
Recently, several extended observation campaigns on the prominent BL
Lacs PKS 2155$-$304, Mrk\,501, and Mrk\,421 by ASCA and BeppoSAX,
partly simultaneously with RXTE and TeV telescopes, have revealed that
in general the X-ray spectral index and the peak energy correlate well
with the source intensity (for a review see Pian 2002).  The emission
of the soft X-rays is generally well correlated with that of the hard
X-rays and lags it by 3$-$4\,ks (\cite{TA96} 1996, 2000, \cite{ZH99}
1999, \cite{MA00} 2000, \cite{KA00} 2000, \cite{FO00} 2000).  However,
significant lags of both signs were detected from several flares
(\cite{TA01} 2001). From \xmm observations of PKS 215-304 \cite{ED01}
(2001) give, however, an upper limit to any time lags of $|\tau|\leq
0.3\,$hr.  They suggest that previous claims of time lags of soft
X-rays with time scales of hours might be an artifact of the periodic
interruptions of the low-Earth orbits of the satellites every $\sim$
1.6 hours.  Large flares with time scales of $\sim$ 1 day were
detected with temporal lags of less than 1.5 hours between X-ray and
TeV energies (for Mrk\,421 see \cite{TA00} 2000).  For all three
sources the structure function and the power density spectrum analysis
indicate a roll-over with a time scale of the order of 1 day or longer
(\cite{KA01} 2001) which seems to be the time scale of the successive
flare events. On shorter time scales only small power in the
variability is found with a steep slope of the power density spectrum
$\sim f^{- (2 - 3)}$ (\cite{TA02} 2002).
  
These results were obtained from data with a relatively low
signal-to-noise ratio, integrated over wide time intervals (typically
one satellite orbit).  Uninterrupted data with high temporal and
spectral resolution can now be provided by \xmm.  From an analysis of
early data taken with the \xmm EPIC cameras from Mrk\,421, the
brightest BL Lac object at X-ray and UV wavelengths, for the first
time the evolution of intensity variations could be resolved on time
scales of $\sim 100\,$s (\cite{BR01} 2001).  Temporal variations
by a factor of three at highest X-ray energies were accompanied by
complex spectral variations with only a small time lag of $\tau =
265^{+116}_{-102}$ s between the hard and soft photons.

In an extensive study of all currently available \xmm observations of
Mrk\,421 \cite{BR03} (2003) find that the source exhibits a rather
complex and irregular variability pattern - both, temporarily and
spectrally.  In general, an increase in flux is accompanied by a
hardening of the spectrum as expected from a shift of the Synchrotron
peak to higher energies.  But there are exceptions and the rate of the
spectral changes varies strongly.  The shortest variability time
scales appear to be of the order of $\simgt\,$ks.  The lags between
the hard and soft band flux are small and can be of different sign.
  
Correspondingly, it is hard to deduce uniquely the underlying physical
parameters for the radiation process from the observations.  For the
currently favored 'shock-in-jet' model for the BL Lac emission (see,
for example, \cite{SP01} 2001) this implies that we are seeing the
emission from multiple shocks which have either largely different
physical parameters or that we detect the emission from similar shocks
at very different states of their evolution, being additionally masked
by relativistic beaming and time dilatation effects.
 
The internal shock scenario assumes that an intermittently working
central machine produces blobs of matter moving at different
velocities along the jet.  The interaction of two blobs is modeled as
the collision of two shells whose interaction starts from the time of
collision (\cite{SI01} 2001; Spada et al. \cite{SP01} 2001;
\cite{MO03} 2003; \cite{TA03} 2003). This time is estimated from the
relative velocity of two shells. During the interaction an internal
shock propagates through the slower shell and accelerates electrons
which produce the observed radiation (\cite{SP01} 2001, \cite{BW02}
2002).  These analytic models can be used to constrain the dimensions
and physical properties of the emitting regions, but they cannot take
into account the detailed hydrodynamic evolution of the interacting
shells, nor the influence of the external medium prior to the
interaction. To this end we have simulated the two dimensional
axisymmetric evolution of two dense shells moving at different
collinear velocities through a homogeneous external medium.

In \S\,\ref{nummet} we describe the numerical method we have used to
simulate both the hydrodynamics and the temporal evolution of
non-thermal electrons in the fluid. The shock acceleration process is
modeled by two different approaches which are described in detail in
\S\,\ref{results}.  The results of our study are presented and
discussed in \S\,\ref{discussion}, and the conclusions are given in
\S\,\ref{conclusions}.

\section{Numerical method}
\label{nummet}

 We assume that the dynamics of blazars is dominated by the thermal
(baryonic or cold) matter while their emission is produced by a
non-thermal component, which is in agreement with \cite{SI01}
(2001). This is justified in our model because the number densities of
electrons and protons are equal and, thus, the inertia of the baryons
is much larger than that of the leptons.

We have performed a set of two dimensional axisymmetric simulations
(in cylindrical coordinates $r$ and $z$) of dense shells of matter
moving at different relativistic speeds in the same direction. The
shells collide after some time giving rise to internal shocks, where
part of the internal energy of the thermal fluid is transferred to
relativistic electrons producing the observed synchrotron radiation.

The problem is split into two parts, a thermal and non-thermal one.
The evolution of the thermal or hydrodynamic component of rest mass
density $\rho$, pressure $P$, radial velocity $v_r$ and axial velocity
$v_z$ is simulated by means of a relativistic hydrodynamic code. The
code also includes a set of $N$ additional fluid components tracing
the evolution of non-thermal relativistic electrons at different
energies. The tracer fluids of number density $n_i$ $(i=1,\dots,N)$
are advected by the thermal fluid, and are coupled in energy space by
their radiative energy losses.

In the following subsections we detail the algorithms used for the
simulation of the thermal fluid (\S\,\ref{hydro}), of the relativistic
electrons (\S\,\ref{non-thermal}) and of the coupling between them
(\S\,\ref{coupling}).

\subsection{Hydrodynamics}
\label{hydro}

The equations of relativistic hydrodynamics can be cast in a system of
conservation laws of the form
\begin{equation}
\frac{\partial\bf{U}}{\partial t} + 
    \sum_{k=1}^3 \frac{\partial \bf{F}^i}{\partial x_i} = 0\ ,
\label{conslaw}
\end{equation}
where ${\bf{U}}$ and ${\bf{F}^k}$ are the vector of conserved
variables and the flux vectors, respectively (\eg \cite{MA94} 1994).
In the case of axial symmetry and expressed in cylindrical coordinates
$(r,z)$ Eq.\,\eqref{conslaw} reads
\begin{equation}
\frac{\partial {\bf U}}{\partial t} + 
        \frac{1}{r} \frac{\partial r {\bf F}}{\partial r} +
        \frac{\partial \bf{G}}{\partial z} = \bf{S}\ ,
\label{cylcoords}
\end{equation}
where
\begin{equation}
{\bf U} = [\rho \Gamma,\,
           \rho h \Gamma^2 v_r, \,
           \rho h \Gamma^2 v_z, \,
           \rho h \Gamma^2 - P - \rho \Gamma, \,
           \rho {\bf X} \Gamma ]^T  
\label{Ucyl}
\end{equation}
is the vector of conserved quantities and 
\begin{equation}
{\bf F} = [\rho \Gamma v_r, \,
           \rho h \Gamma^2 v_r^2 + P, \,
           \rho h \Gamma^2 v_z v_r, \,
           \rho h \Gamma^2 v_r - \rho \Gamma v_r,\,
           \rho {\bf X} \Gamma v_r ]^T
\label{Fcyl}
\end{equation}
and
\begin{equation}
{\bf G} = [\rho \Gamma v_z, \,
           \rho h \Gamma^2 v_r v_z, \,
           \rho h \Gamma^2 v_z^2 + P, \,
           \rho h \Gamma^2 v_z - \rho \Gamma v_z,\,
           \rho {\bf X} \Gamma v_z ]^T,
\label{Gcyl}
\end{equation}
are the corresponding flux vectors in $r$ and $z$ direction,
respectively.  Note that for reasons of convenience the speed of
light, which is denoted by $c$ elsewhere, is set equal to one in this
subsection.  Consequently, the Lorentz factor is given by $\Gamma
\equiv (1-v_r^2-v_z^2)^{-1/2}$, and the specific enthalpy by $h=1 +
\varepsilon + P/\rho$, where $\varepsilon$ is the specific internal
energy of the fluid.
\begin{equation}
{\bf S} = [ 0, \frac{P}{r}, 0, 0, {\bf 0} ]^T
\label{Scyl}
\end{equation}
is the source vector expressing the non-conservation of the radial
momentum in cylindrical coordinates, and
\begin{equation}
 {\bf X} = [X_1, \dots, X_N]^T,\,\, 
\label{X}
\end{equation}
is an N-component vector with
\begin{equation}
  X_i = \frac{n_i m_{\rm e}}{\rho}\,, i =1, \dots, N\, , 
\label{Xi}
\end{equation}
where $m_e$ is the electron rest mass, and $X_i$ is the mass fraction
of the tracer species $i$ with respect to the mass of the thermal
fluid.

The conservation laws are integrated with the GENESIS code of
\cite{AL99} (1999). This code, which is exploits the piecewise
parabolic method (\cite{CW84} 1984), was suitably modified to
passively advect a set of non-thermal species along with the main
thermal fluid. We assume that the thermal component is a perfect fluid
obeying an ideal equation of state of the form $P = (\gamma_{\rm ad} -
1) \varepsilon \rho$, where $\gamma_{\rm ad}=4/3$ is the adiabatic
index.

\subsection{Non-Thermal population}
\label{non-thermal}

The non-thermal particle population evolves both in space and
time. Limiting the physical conditions in the fluid in such a way that
during a time step non-thermal particles are contained in a single
numerical cell, it is possible to split the evolution of the
non-thermal particles in space and in time. This implies that there is
a minimum magnetic field or, equivalently, a maximum allowed Larmor
radius that depends on the numerical resolution employed (see below).
We point out, however, that there are other approaches which treat the
evolution of non-thermal particles by solving the diffusion-convection
equation (\cite{MI01} 2001, \cite{JRE99} 1999).  The spatial evolution of
the non-thermal particles is done by advecting them along with the
fluid and, thus, their macroscopic velocity field corresponds to that
of the thermal fluid (see \S~\ref{hydro}). The treatment of the
temporal evolution is described in this section.

The non-thermal particle population is assumed to be composed of
ultra-relativistic electrons injected at shocks. Its temporal evolution
is governed by the kinetic equation (e.g., \cite{KA62} 1962):
%
%
\begin{equation}
\frac{\partial n(\gamma,t)}{\partial t} + \frac{\partial}{\partial
\gamma}(\dot{\gamma}n(\gamma,t))=Q(\gamma),
\label{eq:ek}
\end{equation}
where $n(\gamma,t)$ is the number density of electrons having a
Lorentz factor $\gamma$ at a time $t$, and $\dot{\gamma} \equiv
d\gamma / dt$ represents the radiative losses.  In the case of
synchrotron radiation the energy loss in cgs-units is (\eg \cite{RL79}
1979)
%
%
\begin{displaymath}
  \dot{\gamma} = -q B^2\gamma^2 
\end{displaymath}
with
\begin{displaymath}
  q \equiv -\frac{2e^4}{3m_{\rm e}^3 c^5}\ .
\end{displaymath}
Here, $B$ and $e$ are the magnetic field strength and the electron
charge, respectively.

The time independent source term $Q(\gamma)$ present in
Eq.\,\eqref{eq:ek} gives the number of electrons injected at shocks
with Lorentz factor $\gamma$ per unit of time.  Relativistic electrons
are injected in zones of the computational grid which separate shocked
and unshocked thermal fluid. Shocks are detected using the standard
criterion in the Piecewise Parabolic Method (\cite{CW84} 1984)
applied to the thermal fluid. For shock acceleration to take place,
the magnetic field strength has to exceed some minimum value such that
the corresponding Larmor radius $r_{\rm L}$ of the fastest particle is
smaller than the smallest zone size ($\Delta L$). This is consistent
with our splitting of the spatial and temporal evolution of the
non-thermal particles (see above). Therefore, the injection of
electrons at shocks is limited to situations where the inequality
%
%
\begin{equation} 
\frac{r_L}{\Delta L}=\frac{m_{\rm e} c^2}{e B}\frac{\sqrt{\gamma_{\rm
  max}^2 -1}}{\Delta L}\leq \xi\
\label{larmor1} 
\end{equation} 
holds.  Here $\xi$ is a free parameter such that $\xi<<1$ (in our
simulations we take $\xi = 10^{-1}$), and $\gamma_{\rm max}$ is the
maximum Lorentz factor of the particles which are injected into the
zone (see below). In terms of the magnetic field strength, condition
(\ref{larmor1}) becomes
%
%
\begin{equation} 
B\geq \frac{m_e c^2}{\xi e\Delta L}\sqrt{\gamma_{\rm max}^2-1}\ , 
\end{equation} 
or numerically
%
%
\begin{equation} 
B\geq \left(\frac{1.7\cdot 10^3 \rm{cm}}{\xi \Delta 
    L}\right)\sqrt{\gamma_{\rm max}^2-1}\,{\rm G}\ . 
\label{larmor2} 
\end{equation} 

The injected electrons are assumed to have a power-law distribution in
the interval $[\gamma_{\rm min}, \gamma_{\rm max}]$ with a power law
index $p_{\rm inj}$. Then the appropriate time-independent source term
reads
%
%
\begin{equation} 
Q(\gamma)=Q_{0}\gamma^{-p_{\rm inj}} {\cal S}(\gamma;\gamma_{\rm 
  min},\gamma_{\rm max})\ , 
\label{eq:pwlsource} 
\end{equation} 
where ${\cal S}(x;a,b)$ is the interval function:
%
%
\begin{displaymath} 
{\cal S}(x;a,b)=\left\{ \begin{array}{ll} 1 &  
                \mathrm{if}\ a\leq x\leq b \\ 0 & \mathrm{otherwise} 
\end{array} \right. 
\end{displaymath} 

The magnetic field ${\bf B}$ is assumed to be randomly oriented, and
its strength is parametrized by the parameter $\alpha_B$ which is
defined as the ratio between the energy density of the magnetic field
and the thermal energy density of the fluid:
%
%
\begin{displaymath} 
  \frac{B^2}{8\pi} = \alpha_B \frac{p}{\gamma_{\rm ad}-1}. 
\label{eq:alpha_b}
\end{displaymath} 
In our simulations the parameter $\alpha_{\rm B}$ was set equal to a
value of $10^{-3}$ in order to obtain magnetic field strengths of the
order of $0.1\,$G in the emitting region (\cite{BW02} 2002).

The emissivity $j(\nu)$ of the synchrotron radiation for a population
of relativistic electrons with a distribution $n(\gamma)$ is given by
(\eg \cite{RL79} 1979)
%
%
\begin{equation}
 j(\nu) = \frac{\sqrt{3}e^3 B}{4\pi m_{\rm e}c^2} 
          \int_{1}^{\infty}\mathrm{d}\gamma n(\gamma)
          F(\nu/\nu_0\gamma^2)\, , 
\label{eq:synem} 
\end{equation} 
where $F(x)=x\int_{x}^{\infty}\mathrm{d}y\ K_{5/3}(y)$; $K_{5/3}(x)$
is the modified Bessel function of index $5/3$, and
%
%
\begin{displaymath} 
\nu_0 = \frac{3eB}{4\pi m_{\rm e}c} \ . 
\end{displaymath} 

In each zone of the computational grid the non-thermal electron
population is represented by a sum of power laws in $N$ Lorentz factor
intervals
%
%
\begin{equation} 
 n(\gamma, t) = \sum_{i=1}^{N} n_0^i (t) \gamma^{-p_i (t)} 
                {\cal S}(\gamma;\gamma_{i-1},\gamma_i)\ , 
\label{eq:sumpwl} 
\end{equation} 
where $\gamma_{i-1}$ and $\gamma_i$ are the lower and upper boundaries
of the $i$-th power law distribution, which at time $t$ is normalized
to $n_0^i (t)$ and has a power law index $p_i (t)$.  The $\gamma_{i}$
are logarithmically distributed according to
\begin{displaymath} 
  \gamma_i = \gamma_{0} \left(\frac{\gamma_N}{\gamma_0} 
                        \right)^{(i-1)/(N-1)}\ ,\,\,  i=1,\dots,N  
\end{displaymath} 
where $\gamma_0$ and $\gamma_N$ are the lower and upper limit of the
whole energy interval considered for the non-thermal electron
population.

Eq.\,\eqref{eq:ek} can be solved analytically for an initial power law
distribution with no injection, and for a power law injection with no
initial electron distribution (\cite{KA62} 1962). We use these solutions
with a slight modification arising from the fact that we solve the
equation for each energy interval separately.

The solution for the case of an initial power law distribution
\begin{equation} 
  n(\gamma,0) = n_0 \gamma^{-p_0} 
                {\cal S}(\gamma; \gamma_{\rm min}, \gamma_{\rm max}) 
\end{equation} 
of index $p_0$ with no injection ($Q= 0$) is 
%
%
\begin{equation} 
  n(\gamma,t) = n_0 \gamma^{-p_0} (1-q B^2\gamma t)^{p_0-2} 
                    {\cal S}(\gamma/(1-q B^2\gamma t);  
                             \gamma_{\rm min},\gamma_{\rm max})\, . 
\label{eq:snoinj} 
\end{equation} 

When no electrons are initially present ($n(\gamma, 0) = 0$), and when
the injection occurs at a constant rate with a power law of index
$p_c$ ($p_c=2.2$, see \S~\ref{coupling}), \ie
\begin{equation}
  Q(\gamma) = Q_0 \gamma^{-p_c}
              {\cal S}(\gamma; \gamma_{\rm min}, \gamma_{\rm max}) \\
\end{equation}
the solution is given by
%
%
\begin{equation}
  n(\gamma,t) =  \frac{Q_0}{q B^2(p_c-1)}\gamma^{-2} 
                 \left(\gamma_{\rm  low}^{1-p_c} -
                       \gamma_{\rm high}^{1-p_c}
                 \right)
                 {\cal S} (\gamma; 
                           \gamma_{\rm min}/(1+q B^2\gamma_{\rm min}t),
                           \gamma_{\rm max}).
\label{eq:sinj}
\end{equation}

$\gamma_{\rm low}$ and $\gamma_{\rm high}$ are evaluated according to
\begin{displaymath}
 (\gamma_{\rm low},\ \gamma_{\rm high}) 
   = \left\{ \begin{array}{lll}
      (\gamma_{\rm min},\ \gamma_{\rm max}) 
   &;\ \mathrm{if}
   &        \gamma_{\rm max}/(1+q B^2\gamma_{\rm max}t)
       \leq \gamma 
       <    \gamma_{\rm min} \\

      (\gamma_{\rm min},\ \gamma/(1-q B^2\gamma t)) 
   &;\ \mathrm{if}
   &        \gamma_{\rm min}/(1+q B^2\gamma_{\rm min}t)
       \leq \gamma 
       <    \gamma_{\rm min} \\
      (\gamma,\gamma/(1-q B^2\gamma t) 
   &;\ \mathrm{if} &
            \gamma_{\rm min} 
       \leq \gamma 
       \leq \gamma_{\rm max}/(1+q B^2\gamma_{\rm max}t) \\
      (\gamma, \gamma_{\rm max}) 
   &;\ \mathrm{if} &
            \gamma_{\rm max}/(1+q B^2\gamma_{\rm max}t)
       <    \gamma
       \leq \gamma_{\rm max}
\end{array}\right.
\end{displaymath}
The case $\gamma_{\rm low}=\gamma,\ \gamma_{\rm high}=\gamma/(1-q
B^2\gamma t)$ is the solution given by \cite{KA62} (1962). This
solution is also recovered in the limit of an infinite interval
($\gamma_{\rm min}\rightarrow 0$,$\gamma_{\rm max}\rightarrow\infty$).

Because the kinetic equation Eq.\,\eqref{eq:ek} is linear in
$n(\gamma,t)$, it can be solved for the distribution given in
Eq.\,\eqref{eq:sumpwl} using Eq.\,\eqref{eq:snoinj} and
Eq.\,\eqref{eq:sinj} for each term in the sum separately. The new
solution is then again approximated in the form of
Eq.\,\eqref{eq:sumpwl}.

\subsection{Synchrotron radiation}
\label{synchrotron}

The frequency dependent synchrotron emission is computed by
substituting $n(\gamma)$ from Eq.\,\eqref{eq:sumpwl} into
Eq.\,\eqref{eq:synem}. Since synchrotron self-absorption is not
significant in the frequency range considered ($10^{16}$-$10^{19}$
Hz), we compute the observed radiation by summing up the contributions
from all zones of the computational grid in each time step taking into
account the light travel time to the observer.

\subsection{Coupling of the thermal and non-thermal components}
\label{coupling}

In this section we explain how the energy losses of the non-thermal
particles (see \S\,\ref{synchrotron}) are coupled to the thermal
fluid.  There are different ways in which $Q_0$ can be computed from
the macroscopic hydrodynamic quantities. It is important to point out
that the acceleration time scale of the electrons (\cite{BO96} 1996)
is much smaller than the hydrodynamic time scale. Therefore, following
the arguments of \cite{JRE99} (1999), we do not treat the shock
acceleration process microscopically, but instead provide macroscopic
models which describe the effects of the electron injection averaged
over a hydrodynamic time step. Such models are not unique, and they
depend on a number of free parameters, \ie they may produce different
light curves from the same hydrodynamic evolution. We will profit from
this lack of uniqueness by comparing our synthetic light curves with
actual observations. This will allow us to disregard injection models
which do not match observed data. In the following two subsections we
consider two models of electron acceleration, each with different
choices of free parameters.

\subsubsection{Injection model of type-E}
\label{dissmodel}
Once a shock is detected, we compute $\dot{\epsilon}$, the change of
the internal energy of the thermal fluid per unit of time behind
shocks. We assume that $\dot{\epsilon}_{\rm acc}$, the energy density
available per unit of time to accelerate non-thermal electrons, is a
fraction $\alpha_{\rm e}$ of $\dot{\epsilon}$ (\cite{DM98} 1998,
\cite{BM96} 1996), \ie
%
%
\begin{equation}
  \dot{\epsilon}_{\rm acc} = \alpha_{\rm e}  \dot{\epsilon} 
                           = \frac{\alpha_{\rm e}}{\gamma_{\rm ad}-1}\dot{p}\ ,
\label{dissen}
\end{equation}
where $\dot{p}$ denotes the temporal change of the fluid pressure
behind a shock due to the hydrodynamic evolution.  From the definition
of the source term (Eq.\,\ref{eq:pwlsource}) follows
%
%
\begin{equation}
  \dot{\epsilon}_{\rm acc} = \int_{\gamma_{\rm min}}^{\gamma_{\rm max}}
                             \mathrm{d} \gamma\ \gamma m_{\rm e} c^2\
                             Q_0^E \gamma^{-p_{\rm inj}}\ .
\label{accen}
\end{equation}

The back reaction of the energy loss due to particle acceleration on
the flow is incorporated by decreasing the pressure in the zone(s)
where the acceleration takes place during a time interval $\Delta t$.
>From Eq.\,\eqref{dissen} one gets
%
%
\begin{equation}
  \Delta p = - \frac{\gamma_{\rm ad} -1}{\alpha_{\rm e}}
               \dot{\epsilon}_{\rm acc} \Delta t\ ,
\end{equation}
where $\Delta t$ is computed in the rest frame of the zone.

Combining Eqs.\,\eqref{eq:pwlsource}, \eqref{accen} and \eqref{dissen}
one obtains
%
%
\begin{equation}
  Q_0^E = \frac{\alpha_{\rm e} (p_{\rm inj}-2) 
                \gamma_{\rm min}^{p_{\rm inj}-2}
               }{m_{\rm e} c^2 (\gamma_{\rm ad} - 1)
                 (1-\eta^{2-p_{\rm inj}})} \, \dot{p}\ ,
\label{disrc}
\end{equation}
where $\eta \equiv \gamma_{\rm max} / \gamma_{\rm min}$. 

The three free parameters of this injection model are $\alpha_{\rm
e}$, $\eta$ and $\gamma_{\rm min}$, respectively.

\subsubsection{Injection model of type-N}
\label{dindmodel}
This model similar to the previous one in that the source term
normalization $Q_0^{\rm N}$ satisfies the equation \eqref{disrc} but
additionally the number density of electrons $n_e^{\rm acc}$
accelerated within a time step $\Delta t$ is parameterized to be a
fraction $\zeta$ of the number of electrons in the zone
%
%
\begin{equation}
  n_e^{\rm acc}=\zeta\frac{\rho}{m_{\rm p}}\ ,
\label{ndae}
\end{equation}
where $m_p$ is the proton mass. Using Eq.\,\eqref{eq:pwlsource} one
finds
%
%
\begin{equation}
  n_e^{\rm acc} = \Delta t \frac{Q_0^N}{p_{\rm inj}} 
                  \gamma_{\rm min}^{1-p_{\rm inj}}
                  (1 - \eta^{1-p_{\rm inj}})\ .
\label{ndae2}
\end{equation}
Using Eqs.\,\eqref{disrc}, \eqref{ndae} and \eqref{ndae2} one obtains
%
%
\begin{equation}
  \gamma_{\rm min} = \frac{p_{\rm inj} - 1}{p_{\rm inj} - 2}
                     \frac{\alpha_{\rm e}}{\zeta}
                     \frac{1 - \eta^{1 - p_{\rm inj}}}{
                           1 - \eta^{2 - p_{\rm inj}}}
                     \frac{m_{\rm p}}{m_{\rm e} c^2}
                     \frac{\dot{p} \Delta t}{(\gamma_{\rm ad} - 1) \rho}\ ,
\label{dind}
\end{equation}
for the minimum Lorentz factor, and
%
% ...
\begin{equation}
  Q_0^N = \frac{\zeta}{m_{\rm p}}
          \frac{\rho}{\Delta  t}
          \frac{p_{\rm inj} - 1}{
                \gamma_{\rm min}^{1 - p_{\rm inj}} 
                (1 - \eta^{1 - p_{\rm inj}})} \ ,
\label{dindsrc}
\end{equation}
for the source normalization, where $\gamma_{\rm min}$ is given by
Eq.\,\eqref{dind}.

The three free parameters of this injection model are $\alpha_{\rm
e}$, $\eta$, and $\zeta$, respectively.

\section{Results}
\label{results}

\subsection{Hydrodynamic setup}
\label{hydrosetup}
\begin{figure}
\centering
\includegraphics[scale=0.8]{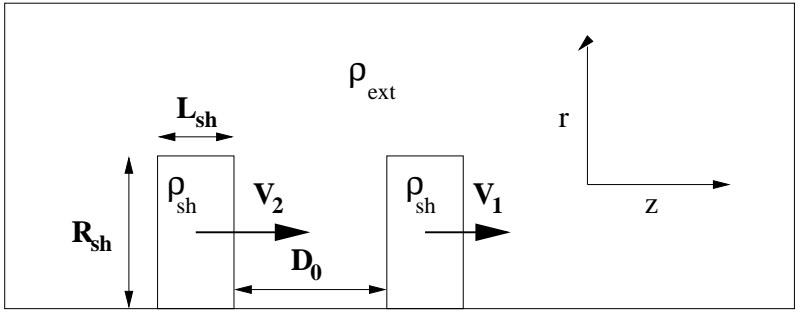}

\caption{A schematic view of the initial setup: Two identical,
cylindrical shells of radius $R_{\rm sh}$, height $L_{\rm sh}$,
density $\rho_{\rm sh}$, and initial separation $D_0$ move through an
external medium of density $\rho_{\rm ext} < \rho_{\rm sh}$ with
velocities $V_1$ and $V_2$ along the symmetry axis ($z$-direction)
such that $V_2 > V_1$. }
\label{hsetup}
\end{figure}

The density of the two colliding, identical shells is $\rho_{\rm sh} =
10^4\,\rho_{\rm ext} = 10^{-22}\,$g cm$^{-3}$, and their temperature
is half the temperature of the external medium $T_{\rm ext} = 7\,
10^7\,$K. The two shells move with Lorentz factors $\Gamma_1=3$ and
$\Gamma_2=15$, respectively. Initially, both shells are of cylindrical
shape with a height $L_{\rm sh} = 10^{14}\,$cm and a radius $R_{\rm
sh} = 10^{16}\,$cm. The shells are initially separated by a distance
$D_0 = 5\,10^{14}\,$cm. The initial setup is shown schematically in
Fig.\,\ref{hsetup}.

The two colliding shells modeled in our simulations are initially
located near the left boundary of the computational grid at $z =
z_{\rm min}$ (\S\,\ref{hsetup}). When the leading of the two shells
approaches the right boundary of the grid at $z = z_{\rm max}$, the
grid is translated into positive $z$-direction such that both shells
are again located near the left boundary of the grid at $z = z_{\rm
min}$.  In order to prevent any numerical artifact due to this
re-mapping close to the left grid boundary at $z_{\rm min}$, we place
the shells sufficiently far from that boundary such that the Riemann
structure emerging from the back of the trailing shell remains
practically unaffected.

The hydrodynamic set up consists of a two-dimensional computational
grid in cylindrical coordinates $(r, z)$ of $40 \times 2000$ zones
covering a physical domain of $1.5\,10^{16}\,$cm by $5\,10^{15}\,$cm.
The grid is initially filled with an external medium at rest, which
has a uniform density $\rho_{\rm ext} = 10^{-26}\,$g\,cm$^{-3}$ and a
uniform pressure $p_{\rm ext} = 10^{-11}\,$erg\,cm$^{-3}$.  After
every re-mapping the computational domain ahead of the shells is
refilled with that external medium. 

We point out that the ratio $\chi = \rho c^2 / 4p$ is rather large
(according to \cite{BW02} 2002) in the shells initially ($\chi \approx
4.5\cdot10^4$), but it decreases considerably during the hydrodynamic
evolution (see next section).

\subsection{Hydrodynamic evolution}
\label{hydroev}
\begin{figure}
\centering
\includegraphics[scale=0.69]{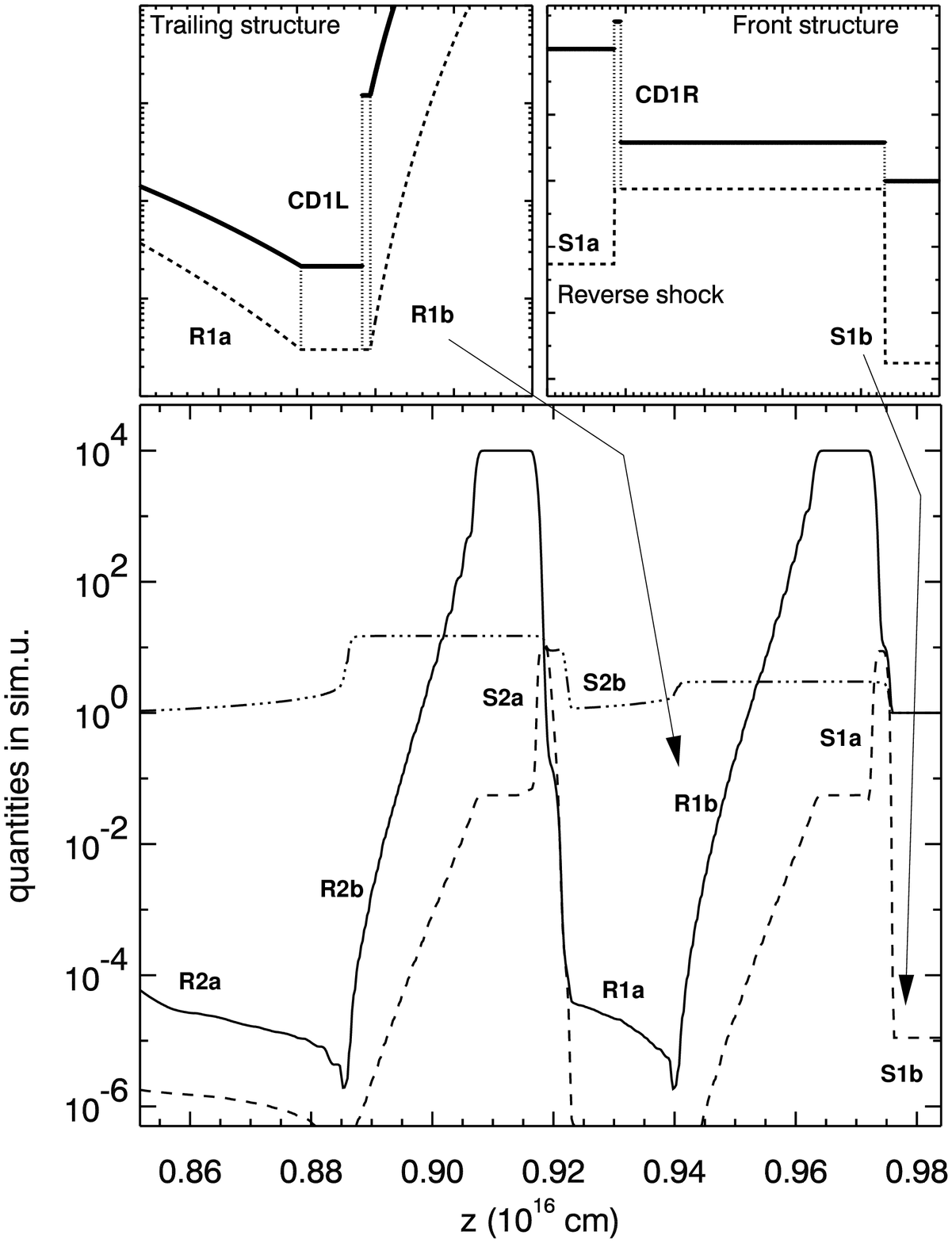}

\caption{Snapshot illustrating the flow structure along the symmetry
axis arising from the setup described in Fig.\,\ref{hsetup} just
before the two shells start to interact ($t=27\,$ks). The lower panel
shows the density (solid line) and pressure (dashed line) distribution
measured in units of $\rho_{\rm ext}$ and $\rho_{\rm ext}c^2$,
respectively. The dash-dotted line gives the Lorentz factor of the
fluid which is moving towards the right. The upper left (right) panel
displays the exact solution of the one dimensional Riemann problem
defined by the trailing (leading) edge of the right shell.  Labeled
are the two bow shocks $S1b$ and $S2b$, the two reverse shocks $S1a$
and $S2a$, the four rarefactions $R1a$, $R1b$, $R2a$ and $R2b$, and
(in the top panels only) the contact discontinuities $CD1L$ and
$CD1R$.}
\label{evscheme}
\end{figure}

As the shells are set up as sharp discontinuities in a uniform
external medium they experience some hydrodynamic evolution before the
actual collision, which starts when the Riemann structure trailing the
slower leading shell meets the bow shock of the fast trailing
shell. This pre-collision evolution is rather similar for both shells,
and can be estimated analytically using an exact one dimensional
Riemann solver. In Fig.\,\ref{evscheme} the two top panels show the
analytic evolution of the flow conditions. The front (with respect to
the direction of motion) discontinuity of each shells decays into a
bow shock ($S1b$ and $S2b$), a contact discontinuity (in
Fig.\,\ref{evscheme} we only show a zoom of the one corresponding to
the leading shell, $CD1R$), and a reverse shock ($S1a$ and $S2a$). The
back discontinuity of each shells develops into a rarefaction ($R1b$
and $R2b$) that connects the still unperturbed state inside the shell
with a contact discontinuity separating shell matter from the external
medium (in Fig.\,\ref{evscheme} only the leading shell is labelled,
$CD1L$), and into a second rarefaction that connects the contact
discontinuity with the external medium ($R1a$ and $R2a$).

The pre-collision evolution is qualitatively similar when instead of
sharp discontinuities a more smooth transition between the shells and
the external medium is assumed. The Riemann structure emerging from
the edges of the shells will be the same, \ie it will consists of the
same structure of shocks and rarefactions as with our set up.
However, the exact values of the state variables in the intermediate
states connecting the conditions in the shells with the external
medium will be obviously different.

The pre-collision hydrodynamics has two direct consequences. Firstly,
each shell is heated by a reverse shock ($S1a$ and $S2a$) which
increases $\mbox{\boldmath$\chi$}$ by \emph{about two orders of
magnitude} (Fig.\,\ref{evscheme}). Secondly, both shells are spread in
$z$ direction as external medium shocked in the bow shocks ($S1b$ and
$S2b$) piles up in front of the shells. The latter effect is
complicated in case of the faster trailing shell by the fact that its
bow shock ($S2b$) soon starts to interact with the rarefaction ($R1a$)
of the slower leading shell. Thereby the bow shock speeds up, and it
eventually catches up with the slower leading shell.  Our simulations
show that the resulting interaction of the two shells occurs at a
distance which is slightly smaller than the distance derived from an
analytic estimate (see below). The accelerating bow shock $S2b$ drags
along the whole Riemann structure. This explains why the state behind
$S2b$ is not uniform (as in case of the slower leading shell), but
shows a monotonically decreasing density and pressure distribution
(Fig.\,\ref{evscheme}). It further explains why the density behind the
reverse shock of the faster shell ($S2a$) is always less than that
behind the reverse shock ($S1a$) of the slower shell.

\begin{figure}
\centering
\includegraphics[scale=0.5]{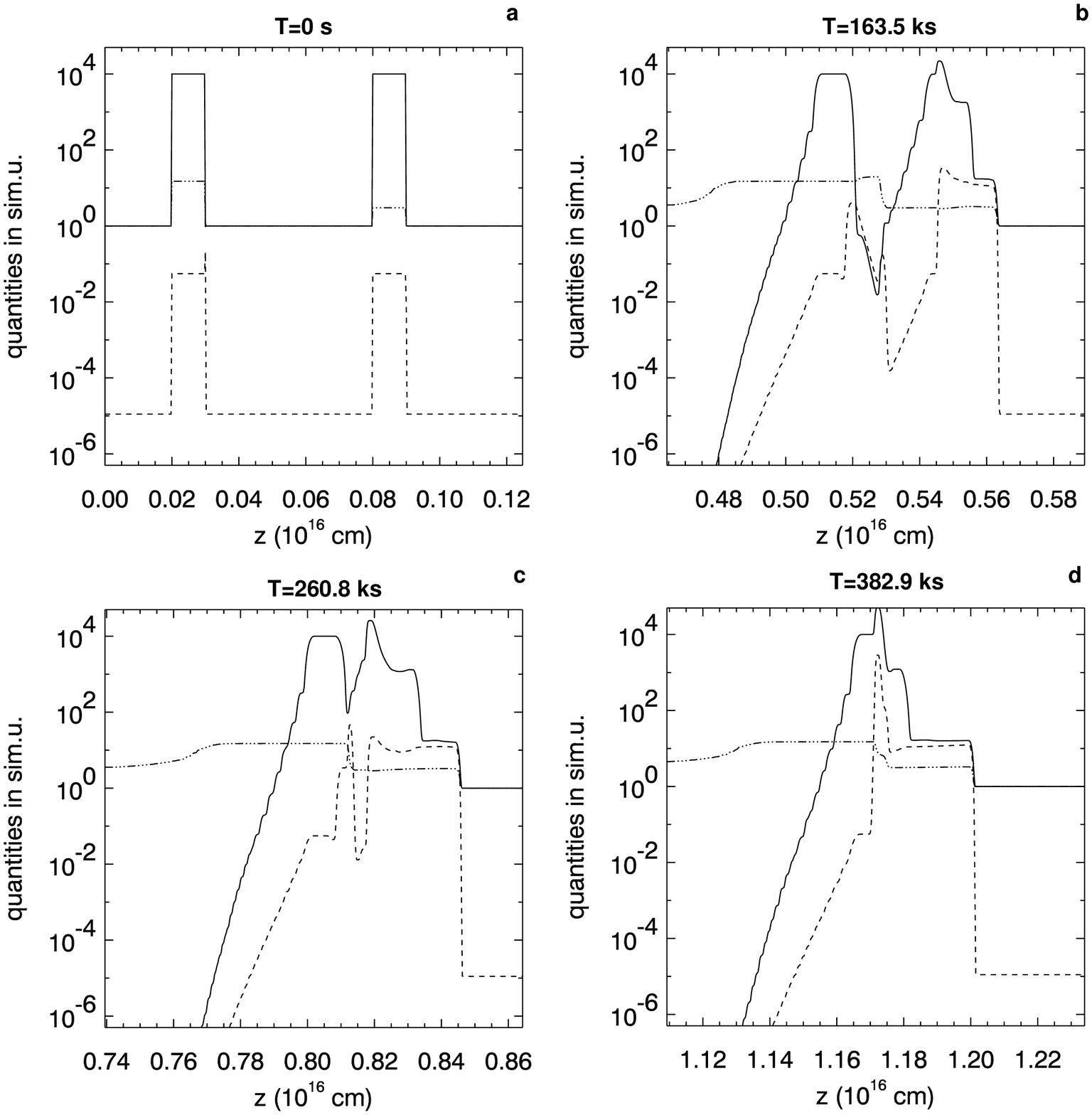}

\caption{Snapshots of the rest mass density (solid line; in units of
$10^{-26}\,$g\,cm$^{-3}$), pressure (dashed line; in units of
$10^{-5}\,$erg\,cm$^{-3}$), and Lorentz factor of the fluid
(dash-dotted line) along the symmetry axis of the computational grid
at four different stages of the evolution.}
\label{1dhydroev}
\end{figure}

\begin{figure}
\centering
\includegraphics[scale=0.57]{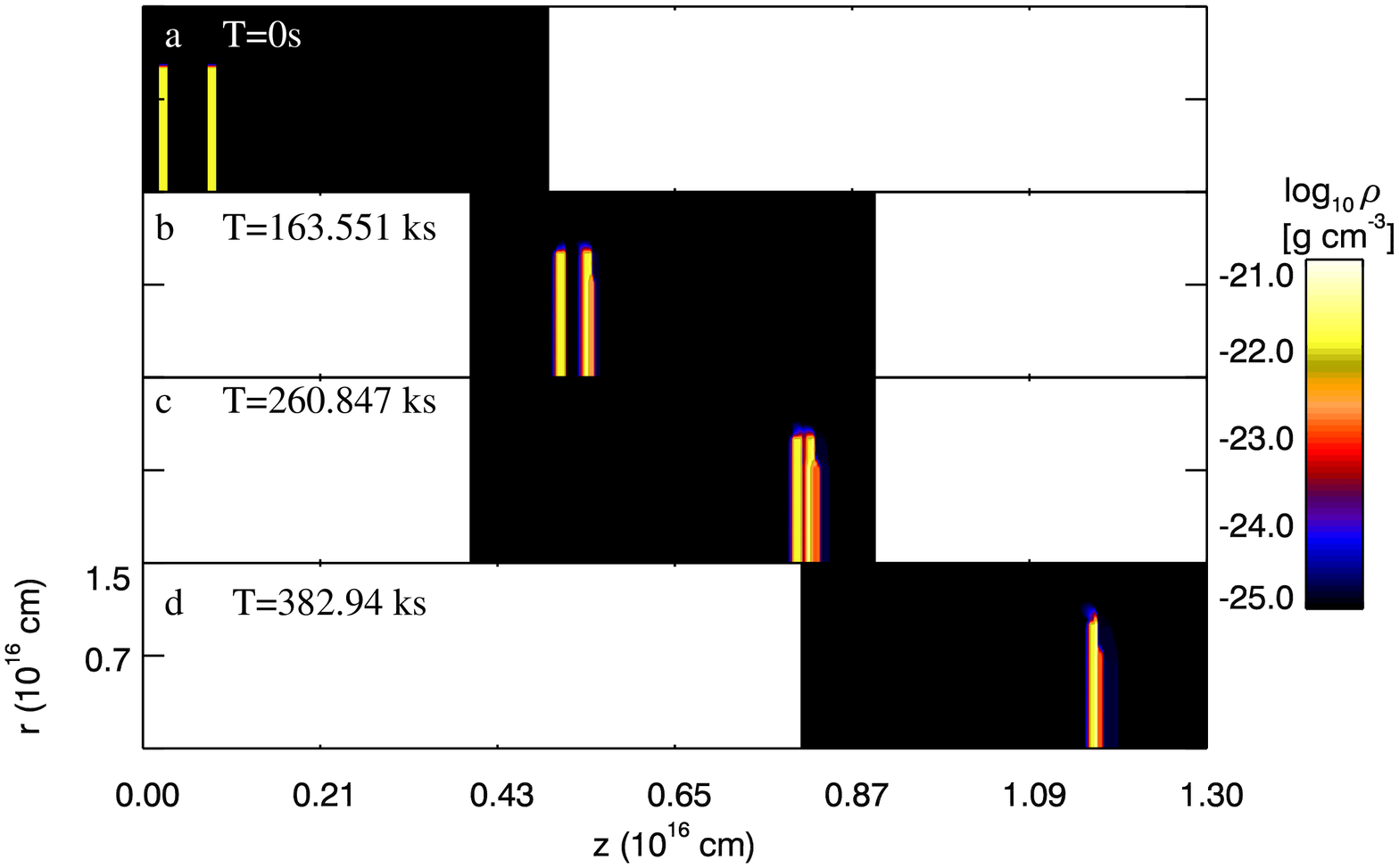}

\caption{Contour plots of the logarithm of the rest mass density at
the same moments of time as in Fig.\,\ref{1dhydroev}. The coordinate
values at both the $r$ and $z$-axis are in units of $10^{16}$\,cm. The
grid re-mapping procedure (see \S\,\ref{hydrosetup}) is the cause for
the shift of the computational domain seen in panels (b) - (d).}
\label{2dhydroev}
\end{figure}

Before the bow shock $S2b$ of the faster trailing shell can enter the
interior of the slower shell where $\rho = \rho_{\rm sh}$, it has to
cross the rarefaction $R1b$, \ie it has to propagate through a
steadily increasing density.  Hence, the emission produced by the
shock will increases gradually during this epoch until it becomes an
internal shock propagating through the slower shell
(Fig.\,\ref{1dhydroev} and \ref{2dhydroev}). We point out that in the
analytic model the internal shock does appear instantaneously when the
two shells touch each other.

Using our initial conditions (see previous section) an analytic
estimate of the time when the two shells collide is given by
%
%
\begin{equation}
 T_c^{\rm an} = \frac{D_0}{V_2 - V_1}
              = \frac{D_0}{c} 
                \left(\sqrt{1-\Gamma_{2}^{-2}} - \sqrt{1-\Gamma_{1}^{-2}}
                \right)^{-1}
              = 280.7\,{\rm ks}\, .
\label{an-est}
\end{equation}
>From our simulation we find that $T_{10} = 0.913\,T_c^{\rm an}$,
$T_{50} = 0.977\,T_c^{\rm an}$ and $T_{90} = 1.009\,T_c^{\rm an}$,
where $T_{10}$, $ T_{50}$ and $T_{90}$ are three moments of time at
which the minimum density ahead of the faster shell is $10\,\%$,
$50\,\%$ and $90\,\%$ of the instantaneous maximum density $\rho_{\rm
max}$ on the faster shell. Note that $\rho_{\rm max}$ does not
coincide, in general, with the initial density of the faster shell
because the shell has undergone some hydrodynamic evolution. Although,
$\rho_{\rm max}$ is only a few per cent larger than $\rho_{\rm sh}$
for the models under consideration.

An analytic estimate of the time at which an observer located at a
distance $ct_0$ from the initial position of the faster shell will
receive the first light is
%
%
\begin{equation}
 T_{\rm arr}^{\rm an} = T_{\rm c}^{\rm an} (1 - V_2/c) + t_0\, .
\label{eq:arrivaltime}
\end{equation}

As mentioned in the previous section, the ratio $\chi$ decreases
during the interaction by almost four orders of magnitude because of
the large pressure increase (from its initial value $\approx 10^4$;
Fig.~\ref{1dhydroev} panel), while the density only increases by a
factor of $\approx 2$. We note that the interaction region is the only
one which produces a significant amount of all observed radiation.
Hence, the value of $\chi$ in that region is the one relevant for
observations and not the initial one.

\subsection{Setup of the non-thermal component}
\label{nsetup}
The non-thermal electrons are binned into $N=48$ logarithmically
spaced intervals covering a total Lorentz factor interval
$[\gamma_1=1, \gamma_N=10^8]$ (\S\,\ref{non-thermal}). The source term
describing the electron injection at shocks (\S\,\ref{coupling}) has a
power law index $p_{\rm inj} = 2.2$, which is compatible with the
value found by Bednarz \& Ostrowski (1998). The injection source terms
are computed for the model type-E (\S\,\ref{dissmodel}) with the
parameters $\alpha_e = 10^{-2}$, $\eta = 10^4$ and $\gamma_{\rm min} =
50$, and for the model type-N (\S\,\ref{dindmodel}) with the
parameters $\alpha_e = 10^{-2}$, $\eta = 10^4$ and $\zeta = 10^{-2}$,
respectively.

In order to produce light curves, the synchrotron emissivity in each
radiating zone is computed in $24$ logarithmically spaced frequency
bands ranging from $10^{16}\,Hz$ to $10^{19}\,Hz$. In a subsequent
step a soft and a hard light curve is obtained by integrating the
computed synchrotron spectra over a prescribed energy interval. For
the soft light curve we considered photon energies between $0.1$ and
$1\,$keV, and for the hard light curve energies between $2$ and
$10\,$keV.

\subsection{Evolution of the non-thermal component}
\label{non-thermalev}

\begin{figure}
\centering
\includegraphics[scale=0.5]{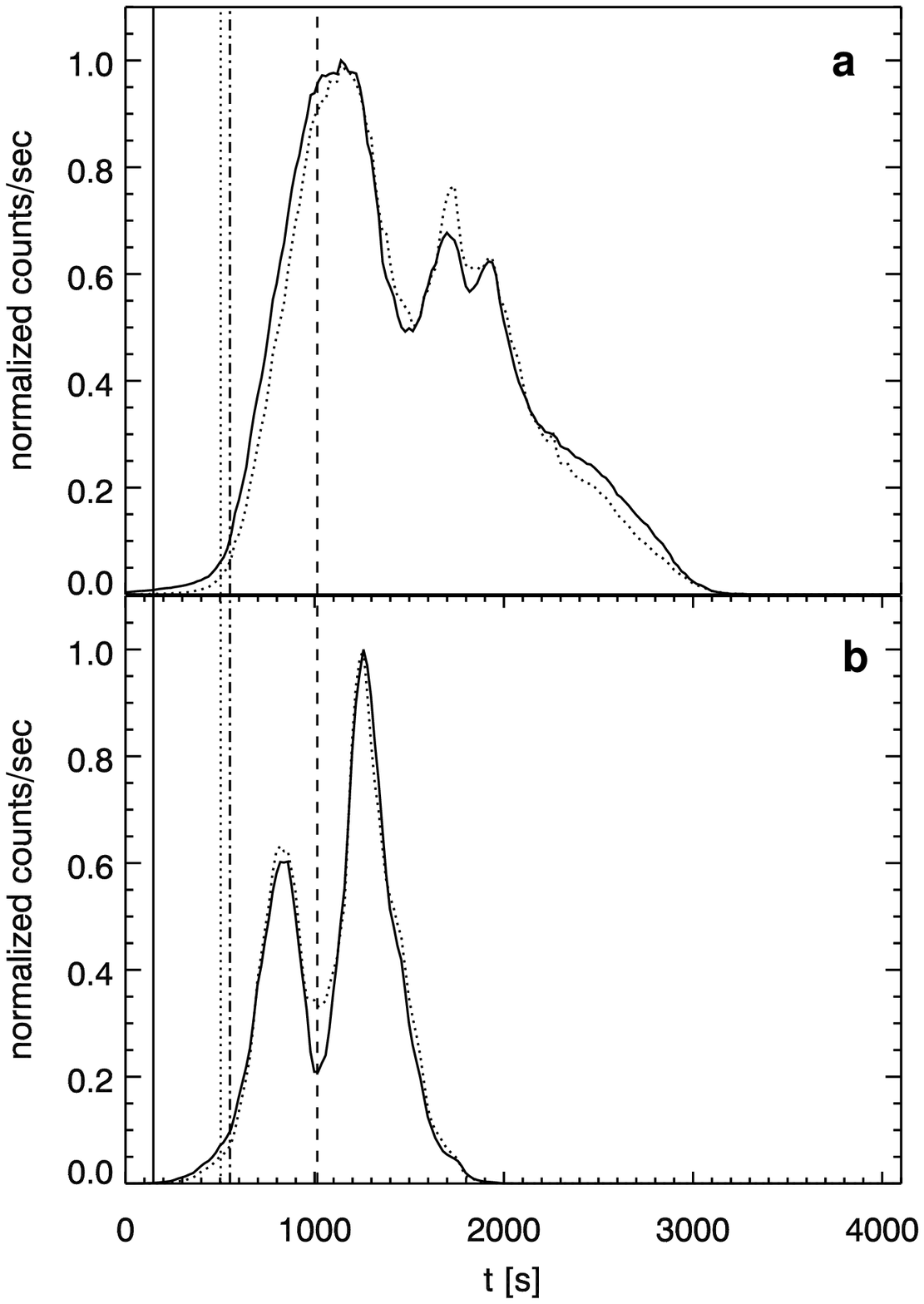}

\caption{Normalized soft (solid line) and hard (dotted line) X-ray
light curves in the observer's frame for injection of type-E (a) and
type-N (b), respectively. The light curves are binned into $20\,$s
time bins.  The vertical lines correspond to $T_{10}$ (solid),
$T_{50}$ (dotted), $T_{90}$ (dashed), and $T_{arr}^{\rm an}$
(dash-dotted), respectively (for a definition of these times see
\S\,\ref{hydroev}).  Note that the time coordinate is renormalized to
the time at which the count rate first exceeds $10^{-3}$ of the count
rate at maximum.}
\label{norm_lc}
\end{figure}

Figure\,\ref{norm_lc} shows the normalized soft and hard X-ray light
curves computed from the hydrodynamic evolution using the procedure
described in the previous subsection. Because the shells start to
interact earlier than predicted by commonly used analytic estimates
(see, \eg \cite{SP01} 2001) the first peak of the light curve is
observed well before the time $T_{90}$ defined at the end of
\S\,\ref{hydroev} (Fig.\,\ref{norm_lc}, panel\,b) for the injection
model of type-N. Such an effect is not present in the injection model
of type-E. The analytically predicted arrival time of the radiation
$T_{arr}^{\rm an}$ (Eq.\,\ref{eq:arrivaltime}) should provide a good
marker for the onset of the light curve. However, $T_{arr}^{\rm an}$
depends on the exact location of the emitting zones and on the
specific injection model used. In the simulated collisions the spatial
arrangement of the emitting zones (shocks) is different from that of
non-evolving shells (analytic case), because some hydrodynamic
evolution already takes place before the shells directly interact.
The arrival time computed from our simulations is hence smaller than
$T_{arr}^{\rm an}$.

\begin{figure}
\centering
\includegraphics[scale=0.65]{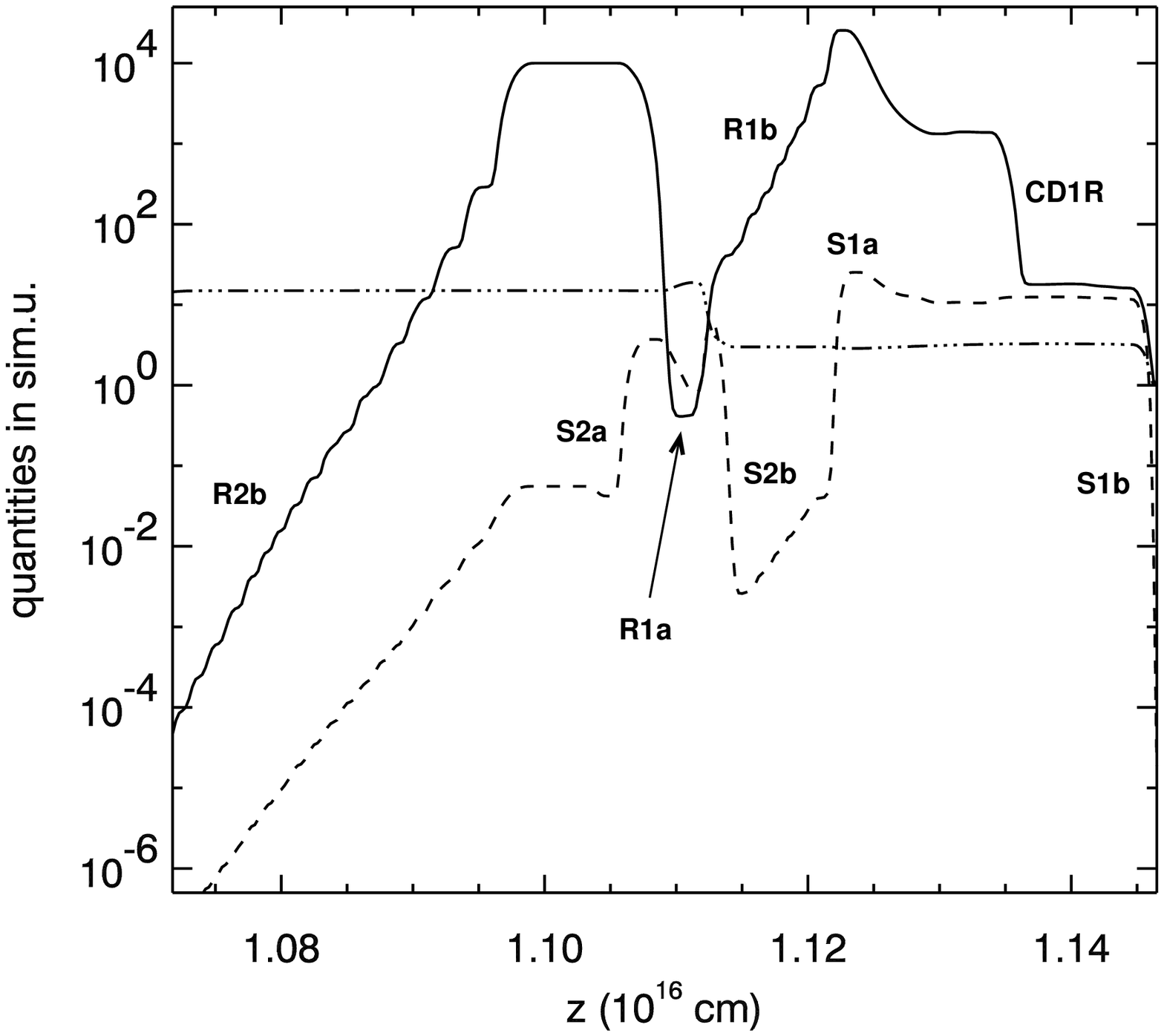}\\

\caption{Snapshot illustrating the flow structure along the symmetry
axis after the two shells start to interact ($t=212\,$ks). The figure
shows the density (solid line) and pressure (dashed line) distribution
measured in units of $\rho_{\rm ext}$ and $\rho_{\rm ext}c^2$,
respectively. The dash-dotted line gives the Lorentz factor of the
fluid which is moving towards the right.  Labeled are the two bow
shocks $S1b$ and $S2b$, the two reverse shocks $S1a$ and $S2a$, the
rarefactions $R1a$, $R1b$ and $R2b$, and the contact discontinuity
$CD1R$. At this time the Riemann structure emerging from the leading
edge of the faster shell is interacting with the trailing Riemann
structure from the rear edge of the slower shell. This leads to the
effective disappearance of the contact discontinuity $CD1L$ and to the
interchange of the position of $R1a$ and $S2b$.}
\label{fig:zoom}
\end{figure}

Our simulations show that the double peak structure of the light
curves is caused by the bow shock $S2b$ propagating into the slower
leading shell, and by the reverse shock $S2a$ propagating through the
faster trailing shell (see Fig.\,\ref{evscheme}). Both of these two
shocks are boosted after the actual shell interaction starts.  We
further find that the relative height of the light curve peaks depends
on the injection model.  In case of the model type-E
(Fig.\,\ref{norm_lc}, panel\,a) the first light curve peak caused by
the (former) bow shock $S2b$ is higher than the second peak caused by
the reverse shock $S2a$, because the rate of change of the energy
density per unit of volume in shock $S2b$ is found to be larger than
in the reverse shock $S2a$. Hence, shock $S2b$ provides a larger
amount of dissipation than shock $S2a$ (see Eq.\,\ref{dissen}), \ie
more electrons are accelerated in the shock $S2b$ than in the shock
$S2a$. In case of the model type-N (Fig.\,\ref{norm_lc}, panel\,b) we
find that less electrons are accelerated in the shock $S2b$ than in
the case of model type-E, because their number density, which is
proportional to the fluid density (Eq.\,\ref{ndae}), is smaller in the
(former) bow shock $S2b$ than in the reverse shock $S2a$ (see
Fig.\ref{fig:zoom}).

Both light curves are binned into $20$s time bins. Thus, if the
temporal resolution of an observations is much worse (effectively it
is $\simeq 80\,$s due to signal-to-noise ratio limitations;
\cite{BR03} 2003), the finest time structures of our computed light
curves will partially or even completely be smeared out, \ie the
differences between the two injection models could not be resolved.

\subsection{Energy}
\label{energy}

Initially, the kinetic ($E_{\rm k}^{\rm ini}$) and thermal ($E_{\rm
th}^{\rm ini}$) energies of the two shells are
\begin{equation}
  E_{\rm k}^{\rm ini}  = L_{\rm sh} R^2_{\rm sh}\pi  \rho_{\rm sh} 
                         (\Gamma_1 + \Gamma_2 - 2) c^2 
                       =  4.4\,10^{46}\,{\rm erg}  
\end{equation}
and
\begin{equation}
  E_{\rm th}^{\rm ini} = L_{\rm sh} R^2_{\rm sh}\pi  
                         \frac{p_{sh}}{\gamma_{\rm ad}-1} 
                       =  4.7\,10^{40}\,{\rm erg}
                       =  9.4\,10^{-7}\,E_{\rm k}^{\rm ini} \, ,
\end{equation}
respectively.  Beyond the time ($t =382.9\,$ks) at which no radiation
is being produced anymore, the kinetic and internal energies of the
fluid are
\begin{equation}
  E_{\rm k}^{\rm fin} = 4.35\,10^{46}\,{\rm erg}
                      = 0.99 E_{\rm k}^{\rm ini}\ ,
\end{equation}
and
\begin{equation}
  E_{\rm th}^{\rm fin} = 4.4\,10^{44}\,{\rm erg}
                       = 0.01 E_{\rm k}^{\rm ini}\ ,
\end{equation}
respectively.  The total radiated energy obtained by integrating the
light curve derived from the integrated synchrotron spectra between
$0.1$ and $10$keV in time is (using model type-N):
\begin{equation}
  E_{\rm rad}^{\rm X} = 1.4\,10^{43}\,{\rm erg} 
                      = 0.03\, E_{\rm Th}^{\rm fin}
\end{equation}
with a peak luminosity of $5.8\,10^{39}\,$erg\,s$^{-1}$. Therefore,
the efficiency of converting thermal energy into X-ray radiation is
$E_{\rm rad}^{\rm X} / E_{\rm th}^{\rm fin} \simeq 0.03$.

We notice that after the collision the resulting merged shell is much
hotter than the initial two shells ($E_{\rm th}^{\rm fin} >> E_{\rm
th}^{\rm ini}$). This heating caused by the internal shocks, and by
the pre-collision hydrodynamic evolution of the shells.

\section{Discussion}
\label{discussion}
Many of the effects found in our axisymmetric simulations can also be
quite accurately modeled assuming a one-dimensional hydrodynamic
evolution. Indeed, the initial set up is explicitly chosen to match
this one-dimensionality, because we want to compare our simulation
results with those obtained from previous analytic one-zone 1D models.
However, we stress that 1D models represents only a first and rough
attempt to model the physics of blazars, which is of course of
multidimensional nature. In the previous sections we have demonstrated
that our simulation results match qualitatively well with those of
previous 1D models. Thus, we are now in the position of exploring
different initial configurations where multidimensional effects are
expected to play a more important role (\eg interaction of shells of
different shell radii, inhomogeneities in the external medium, etc.).

\subsection{Shell interaction}
Concerning the temperature of the shells, which is chosen to be half
the temperature of the external medium (\S\,\ref{hydrosetup}), we do
neither expect any significant change of the dynamic evolution nor the
non-thermal emission, as long as density and pressure contrast between the
shells and the external medium is sufficiently large, and as long as
the temperature within the shells does not greatly exceed that of the
ambient medium. If the latter requirement were to be violated, which
is not expected according to observational data, the shells would
already emit a significant amount of radiation before they collide.

A qualitatively similar shell evolution and shell interaction is to be
expected, if instead of two perfectly (sharp) cylindrical shells one
were to consider shells having a Gaussian (smooth) distribution of
density, pressure, and Lorentz factor. Of course, the exact values of
the arrival time, the time of interaction, etc., would be slightly
different in this case. Decreasing the density, pressure, or Lorentz
factor contrast between the shell and the external medium will
decrease the speed of propagation of the Riemann structures emerging
both from the leading and trailing edges (normal to the direction of
propagation) of the shell. Hence, the collision time would be even
closer to the analytic estimate, and the shells would experience less
pre-collision hydrodynamic evolution (\S\,\ref{hydroev}).

Particularly interesting would be to study (in future work) the
influence of a moving external medium mimicking the flow of an
underlying rarefied jet. In this situation the Lorentz factor contrast
between the shells and the external medium can be substantially
smaller (even close to zero, if the jet moves almost as fast as the
slower shell). Thus, the size and the depth of the Riemann structures
produced by the shells will be much smaller. This has the important
consequence that the effect of the rarefactions trailing the slower
shell on the leading edge of the faster shell will be reduced.

Decreasing the velocity difference between the two shells prolongs the
pre-collision epoch, and reduces the strength of the internal shocks.
Under these circumstances one expects that a smaller amount of energy
is emitted in the X-rays bands.

\subsection{Shock acceleration models} 
The time scale $\tau_{\rm acc}$ of the shock acceleration process
($\tau_{\rm acc} \approx 0.2\,$s for, \eg $B = 0.1\,$G and $\gamma =
5\,10^5$; see \S\,\ref{coupling}) is much shorter than the typical
hydrodynamic time step $\Delta t$ of the simulation ($\Delta t \approx
15\,$s).  Therefore, we have to parameterize the effect of the shock
acceleration process on time scales of the order of the hydrodynamic
time scale.

The two different parameterizations of the shock acceleration process
discussed in \S\,\ref{coupling} produce qualitatively different light
curves. An increase of the fraction of dissipated energy available for
the acceleration of electrons ($\alpha_{\rm e}$) produces either more
electrons (injection model type-E; \S\,\ref{dissmodel}), or injects
electrons at higher energies (injection model type-N;
\S\,\ref{dindmodel}). Increasing the ratio between the upper and lower
Lorentz factor limit of the injection interval ($\eta \equiv
\gamma_{\rm max} / \gamma_{\rm min}$) produces more electrons at
higher energies. Our particular choice of parameters is constrained by
the fact that the maximum of the X-ray synchrotron emission is assumed
to occur around $10^{16}$ - $10^{17}\,$Hz. The particular choice of
the parameter $\zeta$ (the fraction of electrons accelerated in a zone
within a hydrodynamic time step; \S\,\ref{dindmodel}) is suggested by
fits of observed blazar spectra which imply electron number densities
in the range of $10^3 - 10^4\,$cm$^{-3}$ (\cite{MA03} 2003).

\subsection{Efficiency}
The particular shell collision which we have simulated has an
efficiency of the order of $1\,$\% in converting thermal energy to
X-ray radiation (see \S\,\ref{energy}). However, if the flow contains
more than two shells at the same time, the total efficiency of
converting thermal to radiation energy might be larger.

\subsection{Light Curves}
The typical time scale of the flare in the observer's frame is found
to be of the order of one kilosecond (Fig.\,\ref{norm_lc}). However
this finding could be influenced by our particular choice of the shell
width and the magnetic field strength.  The shape of the light curve
will depend on the relation between the time scales associated with
the radiation processes and the physical size of the shells
(\cite{MK02} 2002).  On the one hand, the synchrotron cooling time
scale $\tau_{\rm c}$ of the non-thermal electrons in the comoving
frame of the shell is $\tau_{\rm c} = (qB^2 \gamma)^{-1} =
1.6\,10^4\,$s for $B = 0.25\,$G and $\gamma =5\,10^5$ (see
\S\,\ref{non-thermal}). On the other hand, this time scale is
comparable to the light crossing time of the shell in the same frame
($T_{\rm LC} = L_{\rm sh}/c \Gamma \simeq 3.4\,10^4\,$s; see
\S\,\ref{hydrosetup}). Thus, electrons have sufficient time to radiate
away a large fraction of their energy before they leave the shell.

The light curves show a fast-rise slow-decline structure in both the
soft and the hard energy band (Fig.\,\ref{norm_lc}). It can be
recognized that in the shell collision considered in our simulation no
significant time delays are observed between the hard and soft X-rays.

We have checked that the patterns displayed by, at least, the hard
light curves (Fig.\,\ref{norm_lc}) are similar to the ones found by
\cite{BR03} (2003) for Mrk\,421 in their Figs.\,6 and 9
(corresponding to different orbits of $XMM$), respectively. We point
out that the time scales of the observed light curves of Mrk\,421 are
several times longer than ours. However, this does not invalidate our
comparison, because the duration of the synthetic light curves depends
on the exact physical size of the shells (the wider they are in
$z$-direction, the longer the duration).

\section{Conclusions}
\label{conclusions}
We have performed relativistic hydrodynamic simulations of dense
shells of plasma moving in a rarefied medium, and computed the X-ray
light curves produced during their collision. Our simulations improve
existing analytic (one-zone) models (\eg \cite{SP01} 2001) by
computing a more realistic hydrodynamic evolution of the fluid.

The physical conditions in the external medium correspond to the
standard jet conditions in blazars (\cite{TA00} 2000), while those of
the shells, in particular the density contrast between the shells and
the external medium, are in agreement with the estimates of
\cite{BW02} (2002). These two facts enable us to properly address the
physics of internal shocks in blazars (in contrast to what is claimed
in \cite{BW02} 2002). Indeed, we have demonstrated the ability of our
method to produce synthetic light curves from relativistic
hydrodynamic simulations including the back reaction of the emitted
radiation on the thermal fluid. We find that both the total radiated
energy and the light curves agree well with observational data. With
our method the total efficiency of conversion of thermal energy into
radiation energy needs not to be assumed, but can be directly
computed. For the physical parameters used in our simulations we
obtain a total efficiency of about $1\,$\%.

Our results show that the detailed structure of the synthetic light
curves depends on the particular choice of the macroscopic
parameterization of the process of particle acceleration. By comparing
our results with observed light curves of Mrk\,421 we find that both
models of particle acceleration used in our simulations provide light
curve patterns which can be identified in observed light curves.
However, it seems that our injection model type-E (where a prescribed
fraction of the change of the internal energy of the thermal fluid per
unit time is used to accelerate non-thermal electrons) fits both the
soft and the hard light curves equally well, while the injection model
type-N (similar to type-E, but where only a prescribed fraction of the
available electrons is accelerated) fits only the hard part of the
light curve. The different results obtained with the two models of
injection are related to the fact that while the injection model
type-E accounts for the strength of the shocks (the number of
particles injected is proportional to the rate of change of energy per
unit of volume behind the shock), the model type-N does not.

The light curve produced by a single collision of two shells does not
necessarily produce a single bumped flare, but can consist of several
peaks as our simulations demonstrate. These peaks result from shocks
which form after the actual interaction starts. One of these shocks
propagates into the slower leading shell, while the second shock
propagates into the faster trailing shell. This behavior is clearly
different from that of one-zone models, which only predict a single
peak from a single interaction. A comparison of our results with
observed light curves of Mrk\,421 shows that some of the observed
flares display a multiple peak structure that can be well accommodated
by our model. As a further consequence we note that our model can
explain variability in the light curves with less components (shells)
than one-zone models.

The results presented in this paper represent the beginning of a set
of studies where we will consider different initial configurations in
order to study the influence of such different initial conditions on
the light curves of blazars. Among the possibilities arising we can
immediately include more shells in the simulations, or sticking to
two shell interactions we can vary the size of the shells. The latter
variation, particularly, will allow us to include multidimensional
effects by considering shells of different radial size. A further
improvement of the method will be to include inverse Compton processes
in the treatment of the radiation, which is of particular relevance in
the study of blazars.

The present approach relies on the assumption that the magnetic field
is dynamically negligible, randomly oriented, and its energy density
proportional to the thermal pressure. Thus, another step forward is to
include dynamically important magnetic fields in the simulations.
Finally, we also consider the possibility of performing 3D simulations
which are better adapted to the configurations expected in nature (\eg
non-aligned shells, multidimensional shell trajectories, genuine 3D
shell shapes, etc.). In fact both GENESIS and the newly coded
radiative part are written (and tested) for 3D applications. However,
due to the huge amount of computational power required for a 3D
simulation, we have not yet performed such simulations yet.

\vskip 0.4cm
\begin{acknowledgements}
MAA acknowledges the EU-Commission for a fellowship (MCFI-2000-00504).
PM acknowledges support by the International Max-Planck Research
School on Astrophysics (IMPRS). We are very grateful to Francesco
Miniati for the careful reading of this manuscript and his useful
comments.
\end{acknowledgements}


\begin{thebibliography}{}
%
\bibitem[Aloy \etal]{AL99} Aloy M. A., Ib\'a\~nez, J. M.,
Mart\'{\i}, J. M., \& M\"uller, E. 1999, \apjs, 122, 151

\bibitem[Bednarz \& Ostrowski]{BO98}Bednarz, J. \& Ostrowski, M. 1998,
 Phys.Rev.Lett., 80, 3911

\bibitem[Bednarz \& Ostrowski]{BO96}Bednarz, J. \& Ostrowski, M. 1996,
MNRAS, 283, 447

\bibitem[Bicknell \& Wagner]{BW02} Bicknell, G. V. \& Wagner, S. J., 2002 PASA, 19, 192

\bibitem[Bregman \etal]{BMU87}
 Bregman, J., Maraschi, L.,  \& Urry, C.M. 1987, in {\it Exploring
 the Universe with the IUE Satellite}, ed. Y. Kondo
 (Dordrecht: Reidel),  p. 685

\bibitem[Brinkmann \etal]{BR01} Brinkmann W., Sembay S., Griffiths R.G., \etal 2001,
   A\&A  365, L162

\bibitem[Brinkmann \etal]{BR03} Brinkmann W., Papadakis, I.E., 
  den Herder, J.W.A., Haberl F., 2003, A\&A 402, 929

\bibitem[Bykov \& Meszaros]{BM96} Bykov, A.M. \& Meszaros, P., 1996,
  ApJ 461, L37

\bibitem[Colella \& Woodward]{CW84} Colella, P. \& Woodward, P.R.,
  1984 J.Compt. Phys. 54, 174

\bibitem[Daigne \& Mochkovitch]{DM98} Daigne, F. \& Mochkovitch, R., 1998, MNRAS 196, 275

\bibitem[Edelson \etal]{ED01} Edelson, R., Griffiths, G., Markowitz, A., \etal 2001,
  ApJ, 554, 274

\bibitem[Fossati \etal]{FO00} Fossati G., Celotti A., Chiaberge M., \etal 2000,
  ApJ 541, 153

\bibitem[Jones \etal]{JRE99} Jones, T.W., Ryu, D., Engel, A., 1999 ApJ 512, 105

\bibitem[Kardashev]{KA62} Kardashev, N.S. 1962, Sov. Astr. 6,
317

\bibitem[Kataoka \etal]{KA00} Kataoka J., Takahasi T., 
  Makino F., \etal 2000, ApJ 528, 243

\bibitem[Kataoka \etal]{KA01} Kataoka J., Takahashi T., Wagner S.J.,
\etal 2001, ApJ 560, 569

\bibitem[Malizia \etal]{MA00} Malizia A., Capalbi M., Fiore F., \etal 2000, MNRAS 312,
123

\bibitem[Maraschi \etal]{MA03} Maraschi, L., Tavecchio, F., 2003, ApJ 593, 667

\bibitem[Marti \etal]{MA94} Mart\'{\i}, J.M., M\"uller, E.,
Ib\'{a}\~{n}ez, J.M., 1994 A\&A 281, L9

\bibitem[Mastichiadis \& Kirl]{MK02} Mastichiadis, A., Kirk, J.G.,
2002, PASA, 19, 138

\bibitem[Miniati]{MI01} Miniati, F., 2001, CPC, 141, 17

\bibitem[Moderski \etal]{MO03}Moderski, R., Sikora, M., Blazejowski,
         M., 2003 A\&A, 406, 855

\bibitem[Pian]{PI02} Pian E. 2002, Publ. Astr. Soc. Austr. 19, 49

\bibitem[Sikora \etal]{SI01} Sikora M., Blazejowski, M., Begelman,
M.C., Moderski M., 2001, ApJ, 561, 1154

\bibitem[Rybicki \& Lightman]{RL79} Rybicki, G.B., \& Lightman,
A.P., 1979, Radiative Processes in Astrophyisics, Wiley Interscience
Publication, New York

\bibitem[Spada \etal]{SP01} Spada, M., Ghisellini G.,
Lazzatti, D., \& Celotti, A. 2001, MNRAS 325, 1559

\bibitem[Takahashi \etal]{TA96} Takahashi T., Tashiro M., Madejski G.,
\etal 1996, ApJ 470, L89

\bibitem[Takahashi \etal]{TA00} Takahashi T., Kataoka J., Madejski G.,
\etal 2000, ApJ 542, L105

\bibitem[Tanihata]{TA02} Tanihata C. 2002, PhD Thesis Tokyo Univ., 
  ISAS Research Note 739

\bibitem[Tanihata \etal]{TA03} Tanihata C., Takahashi T., Kataoka J.,
Madejski G. M., 2003, ApJ, 584, 153

\bibitem[Tanihata \etal]{TA01} Tanihata C., Urry C.M. \& Takahashi T.,
\etal 2001, ApJ 563, 569


\bibitem[Urry \& Padovani] {UP95} Urry, C.M., \& Padovani, P. 1995,
PASP, 107, 803

\bibitem[Zhang \etal]{ZH99} Zhang Y.H., Celotti A., Treves A., \etal
1999, ApJ 527, 719

\end{thebibliography}
\end{document}